\documentclass[aps,twocolumn,showpacs,preprintnumbers,amsmath,amssymb,prb]{revtex4-1}
\usepackage[pdftex]{graphicx}
\usepackage[utf8]{inputenc}
\usepackage[percent]{overpic}%

\def \drawone#1#2{\centering{\includegraphics*[scale=#2,clip=false]{#1}}}

\DeclareGraphicsExtensions{.jpg, .png , .pdf, .bmp}

\begin{document}

\title{Point defect modeling in materials: coupling ab initio and elasticity approaches}

\author{Céline Varvenne}
\author{Fabien Bruneval}
\author{Mihai-Cosmin Marinica}
\author{Emmanuel Clouet}
\email{emmanuel.clouet@cea.fr}
\thanks{Corresponding author}

\affiliation{CEA, DEN, Service de Recherches de Métallurgie Physique, F-91191 Gif-sur-Yvette, France}

\date{\today}


\begin{abstract}

Modeling point defects at an atomic scale requires to take special care
of the long range atomic relaxations. 
This elastic field can strongly affect point defect properties 
calculated in atomistic simulations, because of the finite size 
of the system under study.
This is an important restriction for ab initio methods 
which are limited to a few hundred atoms. 
We propose an original approach coupling ab initio calculations 
and linear elasticity theory 
to obtain the properties of the isolated point defect
for reduced supercell sizes.
The reliability and benefit of our approach are demonstrated for
three problematic cases: the self-interstitial in zirconium,
clusters of self-interstitials in iron, and the neutral vacancy in silicon. 
\end{abstract}

\maketitle

\section{Introduction}

Point defects in crystalline solids play a crucial role 
in controlling material properties and their kinetic evolution.
This is true both for intrinsic defects, such as vacancies, self-interstitials
and their small clusters, and extrinsic defects, such as impurities and dopants.
As a consequence, a proper understanding and modeling of material properties often require 
a precise knowledge of point defect characteristics, 
in particular their formation and migration energies.
To this end, ab initio calculations based on the Density Functional Theory (DFT) 
have appeared as a valuable tool.
They are now able to predict point defect energetics
from which one can build quantitative models of material macroscopic behaviors.
Such modeling approaches
have been successful in answering a large variety of experimental questions,
ranging from diffusion processes,\cite{Ma2010, Wu2011}
phase transformations,\cite{Hennig2005}
or recovery of irradiated metals \cite{Fu2004, Fu2005} for instance.
They have also allowed one to predict unsuspected structures of defect clusters,
at sizes where experimental evidence is difficult to obtain.\cite{Terentyev2008, Marinica2012,VeyssierePRB2011}

Ab initio calculations of point defects are currently performed with the supercell approach where periodic boundary conditions are applied. The structure and energy of the point defect are obtained after relaxation of the atomic positions, possibly under various constraints. As ab initio methods are technically limited to a few hundred atoms, the question of the interaction of the defect with its periodic images merits some consideration. If long-range interactions are involved, the convergence of the results with the supercell size - and consequently the ability to obtain the properties of isolated defects - can be out of reach.
This problem is well-known for charged defects, where the long-range Coulomb interaction is involved. Corrective approaches 
\cite{Leslie1985,Makov1995,Freysoldt2009,Taylor2011} are now commonly applied to improve the convergence of these charged defects calculations.
But even neutral defects lead to long range interactions because of their elastic field. 
In the case of linear defects such as dislocations, some specific modeling techniques 
have been developed to circumvent this problem and obtain dislocation intrinsic properties.
\cite{Sinclair1978, Woodward2002, Clouet2009b}
But this problem seems to have been overlooked for point defects.

A point defect in a bulk material induces a long-range elastic field: 
the magnitude of the associated displacements decays like $1/R^2$
with $R$ the distance to the defect.
No characteristic length can be associated with such a decrease 
and the properties obtained by ab initio calculations are therefore
those of a periodic arrangement of interacting defects. 
The commonly applied technique to minimize this artifact is simply to increase the supercell size,
but the sizes necessary to obtain reasonably converged values are sometimes too large
to be handled by ab initio calculations.
This is the case for defects leading to strong elastic fields,
like interstitials or small defect clusters, 
or for materials where a complex treatment of electronic interactions
is required (hybrid functionals, GW methods, ...).

In this article, we propose to couple elasticity theory and ab initio calculations to study point defects. 
We use elasticity theory to model the interaction of the point defect with its periodic images
so as to withdraw this interaction from the ab initio calculations 
and thus obtain the properties of the isolated defect.
The benefit of this approach is demonstrated for three different systems, 
the self-interstitial in zirconium, clusters of self-interstitials in iron,
and the vacancy in silicon. 
These systems differ not only by the nature and the size of the point defect 
but also by the character of the chemical bonding, either metallic or covalent,
and the structure of the crystal, either hexagonal-closed-packed (hcp),
body-centered-cubic (bcc), or diamond.
In all cases, our coupling approach improves the convergence 
with respect to the supercell size, thus allowing a more accurate description 
of point defects than could be achieved with a simple ab initio calculation. 


\section{Modeling approach}

Let us consider a supercell with fixed periodicity vectors $\mathbf{A}_1$, $\mathbf{A}_2$ and $\mathbf{A}_3$
containing one point defect.
After relaxation of the atomic positions, the energy of the supercell
as supplied by the ab initio calculation, $E^{\rm D}_{\varepsilon=0}$, is given by: 
\begin{equation}
\label{eq:E_DP}
E^{\rm D}_{\varepsilon=0} = E_{\infty}^{\rm D} + \frac{1}{2} E^{\rm p}_{\rm int},
\end{equation}
where $E_{\infty}^{\rm D}$ is the energy of the isolated defect and $E^{\rm p}_{\rm int}$
the interaction energy of the defect with its periodic images.
The factor $1/2$ arises because one half of this interaction is devoted to the defect itself, 
and the other goes to its periodic images.  
We use continuous linear elasticity theory to evaluate $E^{\rm p}_{\rm int}$.
Within this theory, a point defect can be modeled by an equilibrated distribution of forces, 
\cite{Teodosiu1982, Bacon1980} i.e. a distribution with no net force nor torque.
If we only retain the first moment of this distribution, 
the defect is fully characterized by its elastic dipole $P_{ij}$.  
The interaction energy $E^{\rm p}_{\rm int}$ of Eq.~(\ref{eq:E_DP}) 
is then evaluated by considering the interaction energy of the point defect
with the strain $\varepsilon_{ij}^{\rm p}$ created by its periodic images:
\footnote{Summations over Cartesian repeated indices are implicit}
\begin{eqnarray}
\label{eq:Epint} 
E^{\rm p}_{\rm int} & = & - P_{ij}\varepsilon^{\rm p}_{ij},  \\
\label{eq:eps_p}
\text{with } \quad \varepsilon^{\rm p}_{ij} & = & -{\sum_{n,m,p}}'G_{ik,jl}(\mathbf{R}_{nmp})\,P_{kl}.
\end{eqnarray}
$\mathbf{R}_{nmp}=n\mathbf{A}_1+m\mathbf{A}_2+p\mathbf{A}_3$,  with $n$, $m$ and $p \in \mathbb{Z}$,
corresponds to the position of the defect periodic images 
and the term $n=m=p=0$ has been excluded from the sum in Eq.~(\ref{eq:eps_p}) (no self-interaction term
as indicated by the prime sign). %
$G_{ik,jl}(\mathbf{x})$ is the second derivative of the anisotropic elastic Green's function 
with respect to the Cartesian coordinates $x_j$ and $x_l$.
Knowing the elastic constants $C_{ijkl}$ of the perfect crystal, 
it is calculated with the numerical scheme proposed by Barnett.\cite{Barnett1972b}
Due to the $1/R^3$ decrease of $G_{ik,jl}(\mathbf{R})$, the lattice summation required in Eq.~\ref{eq:eps_p} is conditionally convergent. 
To regularize the summation, we use the procedure introduced by Cai et al.,\cite{Cai2003} which is based 
on the fact that the displacement and strain fields are necessarily periodic with the same periodicity as the supercell.
Therefore, once the dipole tensor $P_{ij}$ is identified, 
the interaction energy $E^{\rm p}_{\rm int}$ of the point defect with its periodic images can be numerically evaluated 
thanks to Eqs.~(\ref{eq:Epint}) and (\ref{eq:eps_p}).

As previously shown in Ref.~\onlinecite{Clouet2008}, 
the elastic dipole $P_{ij}$ can be directly extracted from the atomistic calculations.
It is linked to the homogeneous stress $\sigma_{ij}$
of a periodic simulation cell of volume $V$ containing one point defect 
through the equation:
\begin{equation}
\label{eq:Pij_mes}
P_{ij}=V \left(C_{ijkl}\varepsilon_{kl}-\sigma_{ij}\right),
\end{equation}
where $\varepsilon_{ij}$ is the homogeneous strain applied on the supercell.
In particular, the elastic dipole is proportional to the homogeneous stress 
in the case of atomistic calculations with fixed periodicity vectors
($\varepsilon=0$).
Compared to other methods where the elastic dipole is either obtained 
from a fitting of the displacement fields \cite{Chen2010a} 
or from the calculation of the Kanzaki forces,\cite{SimonelliPRB1994,Domain2001a,Hayward2012}
Eq.~(\ref{eq:Pij_mes}) presents the advantage of being straightforward and simple to use.

To summaries our approach, once point defects energies have been calculated with ab initio methods, 
they are corrected by subtracting $\frac{1}{2} E^{\rm p}_{\rm int}$, the spurious interaction energy 
arising from periodic boundary conditions, to obtain the properties of isolated defects 
(Eq.~(\ref{eq:E_DP})). After correction, these properties are expected to be weakly sensitive to the supercell size and shape.
The evaluation of the interaction energy 
does not involve any fitting procedure, 
but is a fast post-treatment,
which only requires the knowledge of the elastic constants of the perfect crystal
and the residual stress of the supercell containing the defect.


\section{self-Interstitial in hcp zirconium}

\subsection{Formation energy}

\begin{figure}
\drawone{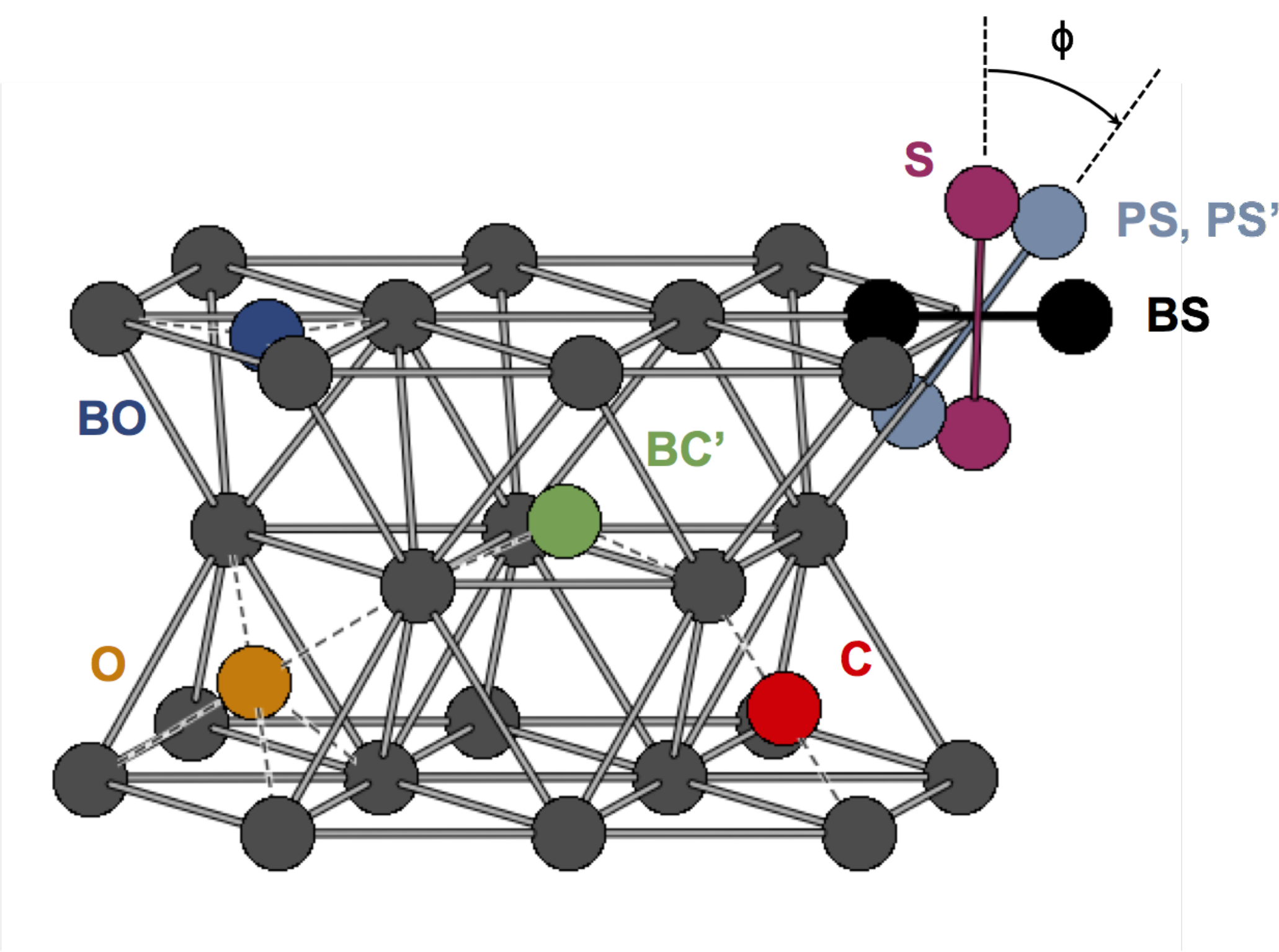}{0.38} 
\caption{Structures of the stable SIA configurations in hcp Zr: 
Octahedral (O), 
basal octahedral (BO),  
split dumbell (S),
basal split dumbell (BS),
crowdion (C) 
and buckled basal crowdion (BC').
PS and PS' are obtained by a rotation of angle $\phi=30^{\circ}$ and $50^{\circ}$ of S
in the prismatic plane.\cite{Verite2013}}
\label{fig:Struct_SIAs}
\end{figure}

We apply this modeling approach to study the self-interstitial atom (SIA) in hcp zirconium.
This point defect appears under irradiation and 
its fast diffusion in the basal planes of the hcp lattice
is often assumed to explain the self-organization of the microstructure
observed in irradiated zirconium,\cite{Woo1988,Samolyuk2013} 
as well as the breakaway growth visible 
for high irradiation doses.\cite{Woo1988,Christien2009}
Recent ab initio calculations \cite{Samolyuk2013,Verite2013}
have enlightened that SIAs in zirconium
can adopt different configurations nearly degenerated in energy.
These configurations are sketched in Fig.~\ref{fig:Struct_SIAs}.
Because of the strong elastic field created by the point defect, 
the associated formation energies
vary with the supercell size, making it hard to get 
a clear view of the SIA energy landscape.\cite{Samolyuk2013,Verite2013}

\begin{figure}
\drawone{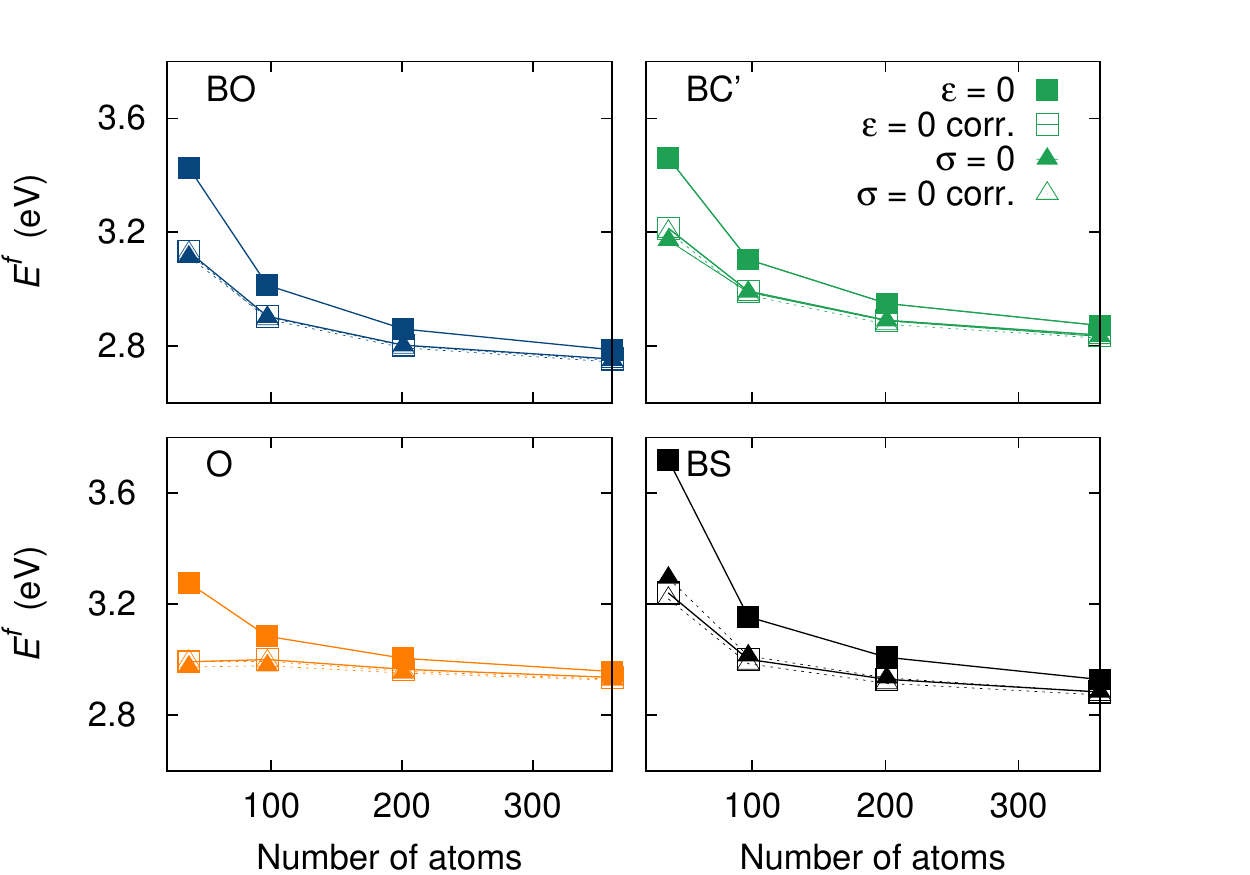}{0.77}
\caption{Formation energies $E^{\rm f}$ of the four most stable SIAs in Zr versus supercell size. 
Filled symbols refer to ab initio uncorrected results 
and open symbols to the results corrected by the elastic model.
The periodicity vectors of the supercell have been either kept fixed (square)
or relaxed (triangles) in the ab initio calculations.}%
\label{fig:Ef_SIAs_corr}
\end{figure}

\begin{figure}
\drawone{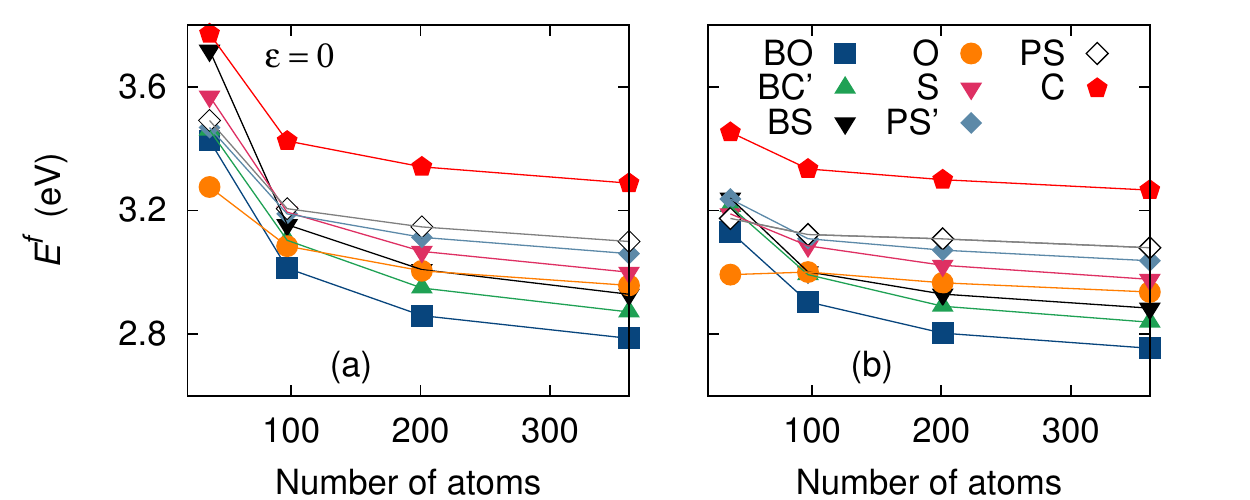}{0.73}
\caption{Uncorrected (a) and corrected (b) Zr SIAs formation energies $E^{\rm f}$ 
of the stable configurations versus the number of atoms for $\varepsilon=0$ calculations.}%
\label{fig:Ef_SIAs_tot}
\end{figure}

We calculate the formation energy of the SIA different configurations 
in the Generalized Gradient Approximation (GGA PBE \cite{Perdew1996}) with an ultrasoft pseudopotential
using the \textsc{Pwscf} code \cite{Giannozzi2009} of the Quantum Espresso package. 
\footnote{Details of the ab initio calculations for Zr can be found
in Ref.~\onlinecite{Clouet2012}.}
Fig.~\ref{fig:Ef_SIAs_corr} shows the variation with the supercell size
of the formation energies for the four most stable configurations:
three high symmetry configurations --- the octahedral (O), 
basal octahedral (BO) and basal split dumbell (BS) ---
and one configuration with a lower symmetry 
that was identified in Ref.~\onlinecite{Verite2013} 
--- the buckled basal crowdion (BC').
Like previous calculations,\cite{Verite2013,Samolyuk2013}
our DFT results, obtained at constant supercell volume and shape ($\varepsilon=0$), 
show that the formation energies strongly depend on the size and shape of the supercell.
In view of these variations, calculations with at least $361$ atoms
are necessary to get converged values.
In addition to this quantitative aspect, 
the SIA properties are not correctly described, 
even qualitatively, if the supercell is too small. 
Indeed, inversions of stability are observed when the supercell size increases
(Fig.~\ref{fig:Ef_SIAs_tot}a).
For instance, the O configuration is more stable than the BS configuration
below 201 atoms, whereas the opposite is true above.

Including now the elastic correction,
we obtain an improved convergence of the formation energies for all configurations
(Fig.~\ref{fig:Ef_SIAs_corr}, $\varepsilon=0$ corr.).
The deviation to the converged values, 
between $120$ and $300$\,meV for uncorrected DFT calculations at $97$ atoms, 
is reduced to the range between $40$ and $150$\,meV when applying the elastic model.
With this correction, the relative stability of the different defects configurations
is well described for a supercell containing no more than 201 atoms
(Fig.~\ref{fig:Ef_SIAs_tot}b). 

Considering now the full energy landscape of the SIA in hcp Zr,
four other stable configurations are found:
a split dumbell (S) along the $c$ axis, 
a crowdion (C),
and two dumbells (PS and PS') resulting from a rigid rotation of S in the prismatic plane.
\cite{Verite2013}
These configurations have a higher energy than the previous ones. 
The elastic correction also helps improving their convergence 
with an energy landscape still correctly described for 201 atoms.

Our approach, coupling ab initio calculations and elasticity theory, therefore allows 
a better picture of SIA energetics for reduced supercell sizes. 
A drift with the size in the formation energies nevertheless remains.
It probably arises from disturbed atomic forces, 
as these forces are also modified by the presence of the periodic images.
As pointed by Puska et al.,\cite{Puska1998} this can disturb the relaxation process
and thus the defect configuration, leading to a variation 
in the formation energies.

\subsection{Zero stress calculations}

Instead of using fixed periodicity vectors in atomistic calculations ($\varepsilon=0$), 
one can also minimize the energy with respect to these vectors 
so as to obtain zero stress ($\sigma=0$) at the end of the relaxation.
Such conditions are sometimes believed to give a better convergence 
than the $\varepsilon=0$ conditions.
As shown by Fig.~\ref{fig:Ef_SIAs_corr}, this is the case for the different configurations
of the SIA in Zr,
but a variation of the formation energy with the supercell size still remains.
Surprisingly, these uncorrected $\sigma=0$ calculations lead to the same energy variations
as the corrected $\varepsilon=0$ calculations.
Before discussing this point, it is worth seeing how the elastic modeling needs to be adapted
in order to add a correction also to these $\sigma=0$ calculations, 
and maybe improve their convergence.

In this $\sigma=0$ case, a homogeneous strain has been applied to the simulation box.
Eq.~\ref{eq:E_DP} therefore needs to be complemented with the energy contribution of this deformation:
\begin{equation}
\label{eq:contrib_def}
\Delta E_{\varepsilon} = \frac{V}{2}C_{ijkl}\varepsilon_{ij}\varepsilon_{kl} - P_{ij}\varepsilon_{ij}.
\end{equation}
We can still use Eq.~(\ref{eq:Pij_mes}) to link the elastic dipole $P_{ij}$ 
with the homogeneous applied strain and the resulting stress.
In the $\sigma=0$ case, the elastic dipole is proportional to the applied strain. 
We obtain that the energy of the supercell 
containing one point defect is given by
\begin{equation}
\label{eq:E_DP_sig0}
E^{\rm D}_{\sigma=0} = E_{\infty}^{\rm D} + \frac{1}{2} E^{\rm p}_{\rm int} - \frac{1}{2V}S_{ijkl}P_{ij}P_{kl},
\end{equation}
where the elastic compliances of the bulk material $S_{ijkl}$ are the inverse tensor
of the elastic constants $C_{ijkl}$.
Eq.~(\ref{eq:E_DP_sig0}) is now used in combination with 
Eqs.~(\ref{eq:Epint}) and (\ref{eq:Pij_mes}),
to extract the energy of the isolated defect, $E_{\infty}^{\rm D}$
from these $\sigma=0$ simulations. 

The corrected formation energies for $\varepsilon=0$ and $\sigma=0$ simulations
are superimposed (Fig.~\ref{fig:Ef_SIAs_corr}).
This shows the validity of our elastic modeling as the corrected formation energies
do not depend on the simulation conditions for a given supercell size.
As noticed before, the uncorrected $\sigma=0$ corrections merge these corrected energies.
This means that the correction applied to the $\sigma=0$ is null:
the spurious interaction energy $1/2 \ E^{\rm p}_{\rm int}$
is compensated by the energy contribution of the homogeneous strain applied to cancel the residual stress
(last term in Eq.~\ref{eq:E_DP_sig0}).
As we will see latter 
this compensation between different energy contributions is specific to SIAs in zirconium.

As a consequence, $\sigma=0$ calculations appear unnecessary. 
For the same result, one can instead perform $\varepsilon=0$ calculations, 
where the periodicity vectors are kept fixed, and then apply the elastic correction.
We highlight the importance of this point, since $\sigma=0$ calculations necessitate 
an increased number of self-consistent field steps. 
Geometry optimizations at $\sigma=0$ are usually a badly preconditioned problem, so
we propose to avoid them systematically.
Moreover, calculations of energy barriers are routinely done with $\varepsilon=0$ conditions,
whereas $\sigma=0$ conditions seem much more complicated.
As we will see below, our correction scheme can also be applied to these barrier calculations,
and then the $\sigma=0$ calculations are made useless.

\subsection{Migration energy}


\begin{figure}
\drawone{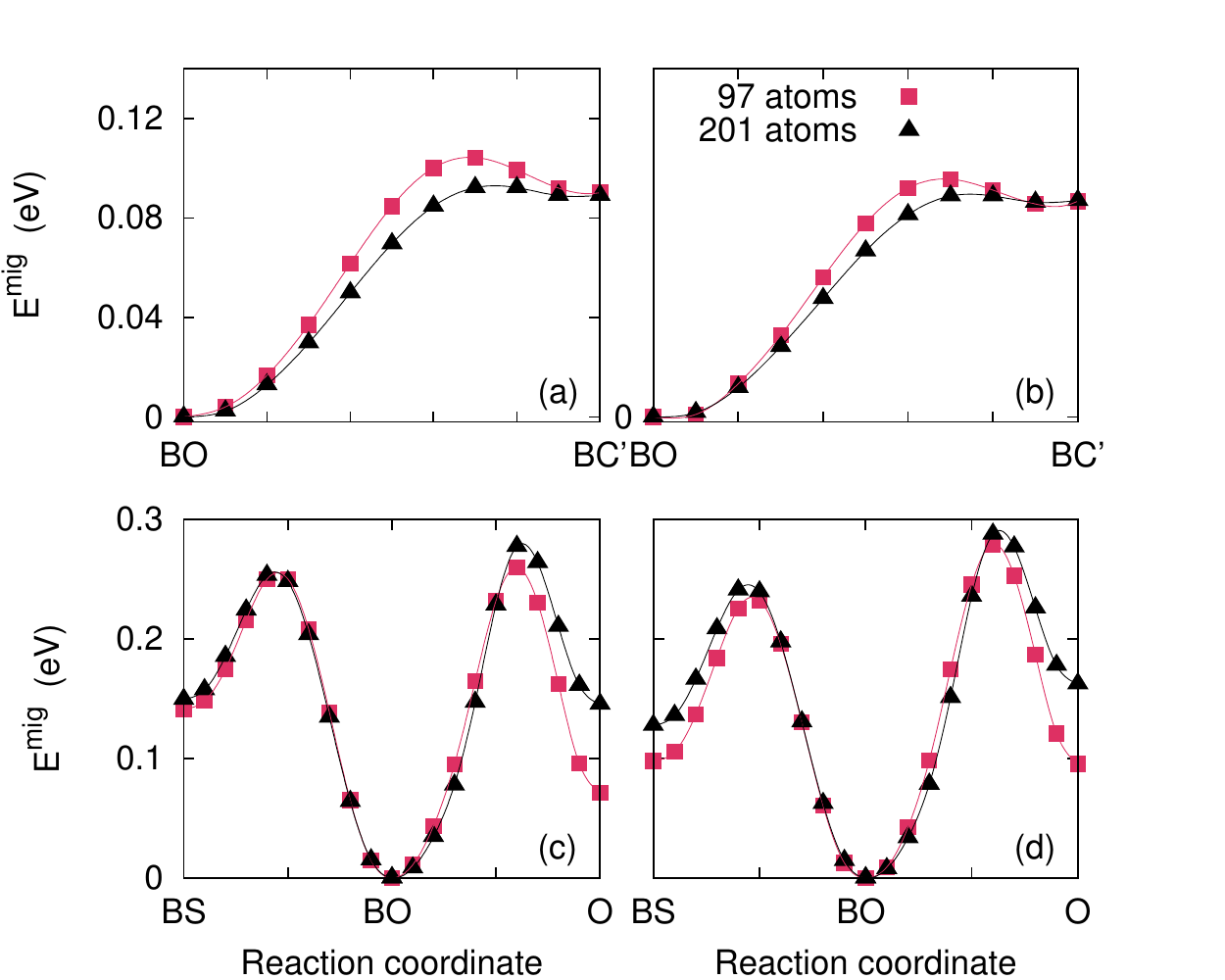}{0.73}
\caption{Migration pathways of Zr SIA calculated with the NEB method, between the BO and BC' configurations and between the BS, BO and O configurations: (a), (c) uncorrected and (b), (d) corrected results.}
\label{fig:Emig_SIAs_corr}
\end{figure}


Our approach, coupling elasticity and ab initio calculations, is not restricted 
to the modeling of stable configurations.
It can also be beneficial to study migration pathways between these configurations.
To illustrate this point, we consider the migration between different configurations 
of the SIA in Zr. 
The minimum energy pathways are investigated using the Nudged Elastic Band method (NEB),\cite{Henkelman2000} 
and the results are presented in Fig.~\ref{fig:Emig_SIAs_corr} for simulation cells containing $97$ and $201$ atoms.

We fist focus on the migration between the two most stable configurations of the SIA in Zr,
namely BO and BC'.
Without the elastic correction (Fig.~\ref{fig:Emig_SIAs_corr}a),
there is a saddle point between these two configurations 
with a supercell containing 97 atoms. This saddle point almost disappears 
with a 201 atom supercell, showing that the transition from BC' to BO is athermal.
Consequently, BC' cannot be considered as a stable configuration:
it corresponds to an extended flat portion of the energy surface with an unstable behavior 
leading to the basin of the BO configuration.
When the elastic correction is included (Fig.~\ref{fig:Emig_SIAs_corr}b),
the result with $97$ atoms already shows a reduced energy barrier,
thus illustrating the acceleration of the convergence with this correction.

We then examine two migration pathways important for the diffusion: 
the transition BO-BS inside the basal plane 
and the transition BO-O along the $c$ axis
Without the elastic correction (Fig.~\ref{fig:Emig_SIAs_corr}c) 
there is no significant difference between these two migration barriers,
even for $201$ atoms NEB calculations. 
On the other hand, the corrected barriers (Fig.~\ref{fig:Emig_SIAs_corr}d)
lead to a migration easier in the basal plane,
with a difference of about $0.07$\,eV in the migration energies.
This could induce a diffusion anisotropy of the SIA at a macroscopic scale.
This of course needs to be confirmed by the calculations of all migration barriers, 
and then the modeling of the diffusion coefficient.

Like for the BO-BC' transition, the elastic correction improves the convergence 
of the BO-O barrier.
But the situation is less clear for the migration from BO to BS.
In this last case, the uncorrected DFT calculation provides indeed superimposed barriers between $97$ and $201$ atoms,
whereas the level of the BS energy changes on the corrected curves.
This can be understood by looking at the formation energies 
of the BO and BS configurations in Fig.~\ref{fig:Ef_SIAs_corr}. 
Without correction, the convergence rate is the same.
There is thus a compensation of errors when considering the energy difference
between these two configurations, and also the migration energy between them.
As a consequence the barriers calculated for 97 and 201 atoms appear superimposed.
Such an error compensation does not occur for the corrected barrier, 
as the convergence rate is not the same for the energies of the BO and BS configurations, 
once corrected (Fig. \ref{fig:Ef_SIAs_corr}).

\section{SIA clusters in bcc iron}

We now look how our modeling approach performs in a case where the point defect 
creates a stronger elastic field than the one of a self interstitial atom (SIA).
To do so, we consider SIA clusters in bcc iron.
SIAs created during irradiation in iron can migrate either to annihilate at sinks 
or to form clusters. 
These clusters adopt different morphologies. 
Large enough clusters have a two-dimensional shape corresponding to dislocation 
loops with a $1/2\,\langle111\rangle$ Burgers vector.\cite{Arakawa2007}
But a broader range of morphologies \cite{Fu2004,Terentyev2008,Marinica2012}
is available to clusters containing a few SIAs.
In particular, it has been shown recently that some clusters 
can have a three-dimensional structure with an underlying crystal symmetry corresponding 
to the C15 Laves phase.\cite{Marinica2012} 
These C15 clusters are predicted to be very stable at small sizes and highly immobile,
in contrast with the $\langle 111 \rangle$ loop clusters which can easily glide 
along the $\langle111\rangle$ direction, leading to a fast 1D diffusion.\cite{Arakawa2007}
Knowing the relative stability of the different configurations that can adopt 
a SIA cluster in iron is of prime importance to be able to model then the kinetic evolution.
The stability of the C15 clusters is closely related to the magnetic properties 
of iron,\cite{Marinica2012,Dezerald2013} which are out of reach of empirical potentials.
Therefore ab initio calculations are needed. 
This severely limits the size of the SIA cluster which can be simulated 
and makes our modeling approach potentially attractive to push back this limit.

\begin{figure}
\drawone{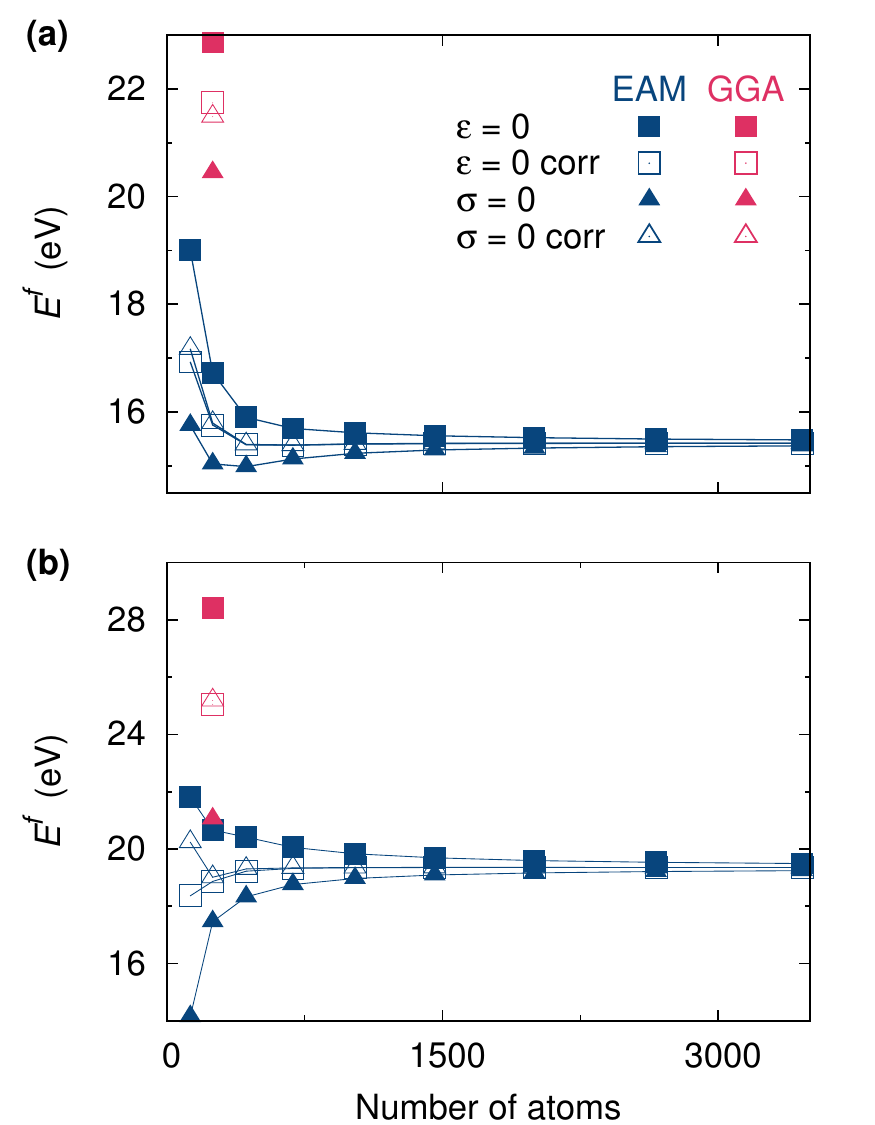}{0.75}
\caption{Formation energy of a SIA cluster containing eight interstitials in bcc iron
calculated for fixed periodicity vectors ($\varepsilon = 0$) or at zero stress ($\sigma=0$)
for different sizes of the simulation cell: 
(a) C15 aggregate and (b) parallel-dumbell configuration with a $\langle111\rangle$ orientation.
Atomistic simulations are performed either with the M07 empirical potential \cite{Marinica2012} (EAM)
or with ab initio calculations (GGA).
Filled symbols refer to uncorrected results and open symbols to the results corrected by the elastic model.}
\label{fig:Ef_8sia111c15}
\end{figure}

To illustrate this point, we consider a cluster containing eight SIAs with two different 
configurations, a C15 aggregate and a planar configuration corresponding to an aggregate 
of parallel-dumbells with a $\langle 111 \rangle$ orientation.
The formation energies of both configurations have been first calculated with the M07
empirical potential \cite{Marinica2012} for different sizes of the simulation cell (Fig. \ref{fig:Ef_8sia111c15}). 
With fixed periodicity vectors of the simulation cell ($\varepsilon=0$), 
one needs at least 2000 atoms for the C15 aggregate and 4000 atoms for the $\langle 111 \rangle$
planar configuration to get a formation energy converged to a precision better than 0.1\,eV.
The convergence is slightly faster for zero stress calculations ($\sigma=0$) 
in the case of the C15 aggregate (Fig. \ref{fig:Ef_8sia111c15}a), 
but the opposite is true in the case of the $\langle 111 \rangle$ planar configuration
(Fig. \ref{fig:Ef_8sia111c15}b).
When we add the elastic correction, the convergence is improved for both cluster configurations.
The corrected $\varepsilon=0$ and $\sigma=0$ calculations lead then to the same formation energies,
except for the smallest simulation cell (128 lattice sites)  
in the case of the $\langle 111 \rangle$ cluster.
This deviation for the smallest supercell is not surprising, 
since the $\langle 111 \rangle$ cluster almost touch its periodic images 
in the simulation cell containing 128 lattice sites. 
In this case, the interaction between the cluster and its periodic images
cannot be reduced only to an elastic interaction.
The problem is not present for C15 clusters which are more compact. 
It is worth pointing that, contrary to the SIA in zirconium,
corrected energies are different and converge faster
than uncorrected energies calculated with the $\sigma=0$ condition.

These formation energies have been also obtained with ab initio calculations
using GGA PBE, a $2 \times 2 \times 2$ k-point grid and an ultrasoft pseudopotential\footnote{Details 
of the ab initio calculations for Fe can be found
in Ref.~\onlinecite{Marinica2012,Dezerald2013}.} 
for a simulation cell containing 250 lattice sites (Fig. \ref{fig:Ef_8sia111c15}).
Calculations with fixed periodicity vectors ($\varepsilon=0$) lead to an energy difference
$\Delta E = -5.6$\,eV between the C15 and the $\langle 111 \rangle$ planar configuration, 
whereas this energy difference is only $\Delta E = -0.6$\,eV in zero stress calculations
($\sigma = 0$). In all cases, the C15 configuration is the most stable but the energy difference 
varies a lot.
Once the elastic correction added, this energy difference is $\Delta E = -3.3$\,eV with 
the $\varepsilon=0$ condition and $\Delta E = -3.7$\,eV with the $\sigma=0$ condition.
Although the size of the simulation cell may appear small compared to the size of the defect, 
a good precision is obtained with this approach coupling ab initio calculations and elasticity
theory. We can conclude that the C15 configuration is the most stable one
with an energy lower by $3.5 \pm 0.2$\,eV than the $\langle 111 \rangle$ planar configuration.

\section{Vacancy in silicon}

\begin{figure}
\drawone{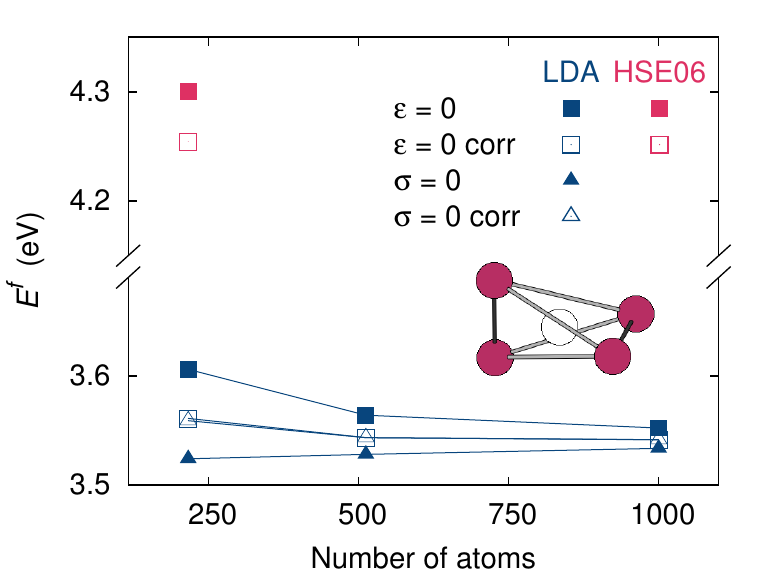}{1.0}
\caption{Vacancy formation energy $E^{\rm f}$ in silicon calculated with the LDA and HSE06 functionals, 
either for fixed periodicity vectors ($\varepsilon=0$) or at zero stress ($\sigma=0$).
Filled symbols refer to the ab initio uncorrected results and open symbols to the results corrected by the elastic model. The vacancy configuration is displayed in the inset: the white sphere corresponds to the empty lattice site and the purple spheres represent its first nearest neighbors.}
\label{fig:Ef_V_Si}
\end{figure}

We finally illustrate the usefulness of our approach by considering another system,
the vacancy in diamond silicon.
This point-defect experiences a strong Jahn-Teller distortion \cite{Watkins1976} 
(see inset in Fig.~\ref{fig:Ef_V_Si}),
leading to a long-range elastic field which disturbs the convergence of ab initio calculations.
To correctly describe the properties of defects in semiconductors, 
one needs to quantitatively predict the size of the band gap.
Simple DFT approximations, like the Local Density Approximation (LDA)
or the GGA,
do not correctly address this problem. 
We have to turn to methods with a higher accuracy, like the random phase approximation
or hybrid functionals.\cite{Bruneval2012}
The  slow convergence of the vacancy formation energy  with respect to the size of the supercell and  the k-point sampling 
\cite{Corsetti2011, Probert2003} then becomes problematic, because the above mentioned 
ab initio methods have a very poor scalability with the system size.

Calculations of the vacancy formation energy within LDA provide a validation of the elastic correction for this defect (Fig.~\ref{fig:Ef_V_Si}): $216$ atom supercells are sufficient to get converged values. 
One can precise that the Jahn-Teller configuration is unstable for smaller systems with LDA.
Once corrected, both $\varepsilon=0$ and $\sigma=0$ calculations lead to the same energies
and converge faster than uncorrected results.
The little remaining drift in the corrected formation energy 
certainly arises from the fact that the tetragonality ratio around the vacancy slightly varies with the supercell size.
As a consequence, we also obtain a small variation of the elastic dipole.
The relaxation process is therefore slightly affected by the presence of the periodic defect images,
leading to the remaining energy variation.

DFT calculations with the hybrid HSE06 functional \cite{Heyd2003, Heyd2006}
stabilize and favor the Jahn-Teller configuration, in agreement with experiments,\cite{Watkins1976} 
but calculations beyond 216 atom supercells are computationally prohibitive.
Note that a fine $2 \times 2 \times 2$ k-point grid was necessary to ensure the appropriate convergence. 
The HSE06 calculation, once corrected, predicts a converged value of $4.26$\,eV,
which is consistent with previously published values.\cite{Gao2013}


\section{Conclusion and perspectives}

In conclusion, we showed in this article that the coupling of ab initio calculations
with an elastic modeling accelerates the convergence of point defect energetics.
The reliability of our approach has been demonstrated on three very different point defects,
a self-interstitial in an hcp metal, a cluster of eight self-interstitials in a bcc metal,
and a vacancy in a diamond semiconductor. 
The corrected results merge the $\sigma=0$ ab initio calculations for the interstitial in zirconium 
but converge faster both for the interstitial clusters in iron and the vacancy in silicon.
This makes useless such $\sigma=0$ calculations.
The elastic correction also applies to energy barriers, 
calculated with the NEB method for instance.

The proposed approach is general and can be directly used for any ab initio study of point defects:
\footnote{The Fortran source code used to calculate the elastic interaction energy 
is provided as supplemental material.} 
once known the elastic constants of the perfect crystal, 
the associated post-processing uses one single piece of information that is anyway calculated in any ab initio code, namely the stress tensor in the defective supercell.
This will make possible the ab initio study of defects for which 
a quantitative description would be out of reach otherwise. 
This includes point defects creating a strong distortion of the host lattice,
large interstitials or small clusters for instance, as well as elements with many electrons, like actinides.
It becomes also conceivable to use ab initio methods giving a more accurate description 
of the electronic structure (all electron methods, hybrid functionals, \ldots),
without a loss of precision induced by the small size of the supercell.

Our elastic correction scheme can also be applied to charged defects, 
where it will sum up with the standard electrostatic correction. 
\cite{Leslie1985,Makov1995,Freysoldt2009,Taylor2011} 
However, the residual stress used as input parameter needs before 
to be corrected from any spurious electrostatic contribution, as discussed in Ref.~\onlinecite{Bruneval2012a}. 

Finally, it is worth pointing out that our approach could be extended 
to correct forces on atoms from disturbances due to periodic boundary conditions.
To do so, one needs to consider the derivative, with respect to atomic positions, 
of the interaction energy appearing in the total energy (Eq. \ref{eq:E_DP}).
With such an elastic correction on the forces, it would be possible 
then to obtain a  better structural relaxation
and to further improve the energy convergence.


\begin{acknowledgments}
  The authors thank J.-P. Crocombette and F. Willaime for fruitful discussions.
  This work was performed using HPC resources from GENCI-CCRT 
  (Grants 2012-096847, 2013-096973, and 2013-096018). 
  AREVA is acknowledged for financial support.
\end{acknowledgments}

\bibliographystyle{apsrev4-1}
\bibliography{correction_elastique}

\end{document}